\documentclass[11pt]{article}
\usepackage{amsfonts}
\usepackage{amssymb}
\usepackage{amsmath}
\usepackage{graphicx}

\def \Zed {\mbox{$\mathbb{Z}$}}

\def \lket {\left|}
\def \rket {\right\rangle}
\def \lbra {\left\langle}
\def \rbra {\right|}
\newcommand{\ket}[1]{\lket #1\rket}
\newcommand{\smallket}[1]{| #1\rangle}
\newcommand{\bra}[1]{\lbra #1\rbra}
\newcommand{\bracket}[2]{\lbra\left. #1\,\right|#2\rket}
\newcommand{\opbracket}[3]{\lbra #1\left|\,#2\,\right|#3\rket}

\def \qed {\hfill \rule{0.2cm}{0.2cm}\vspace{3mm}}
\newenvironment{mylist}[1]
	{\begin{list}{}{\setlength{\leftmargin}{#1}
	\setlength{\rightmargin}{0.0cm}\setlength{\labelsep}{1.3mm}
	\setlength{\labelwidth}{0.8cm}\setlength{\itemsep}{0.0cm}}}
	{\end{list}}
\newtheorem{theorem}{Theorem}
\newtheorem{lemma}[theorem]{Lemma}
\newtheorem{cor}[theorem]{Corollary}

\begin{document}

\title{\Large\bf Quantum simulations of classical random walks\\
and undirected graph connectivity}

\author{John Watrous\thanks{
	Part of this work was performed while the author was at the University
	of Wisconsin--Madison Computer Sciences Department under the support
	of NSF grant CCR-95-10244.}\\
	D\'epartement IRO\\
	Universit\'e de Montr\'eal\\
	Montr\'eal (Qu\'ebec), Canada\\
	watrous@iro.umontreal.ca}

\maketitle

\thispagestyle{empty}

\begin{abstract}
It is not currently known if quantum Turing machines can efficiently
simulate probabilistic computations in the space-bounded case.
In this paper we show that space-bounded quantum Turing machines can
efficiently simulate a limited class of random processes: random walks on
undirected graphs.
By means of such simulations, it is demonstrated that the undirected graph
connectivity problem for regular graphs can be solved by one-sided error
quantum Turing machines that run in logspace and halt absolutely.
It follows that symmetric logspace is contained in the quantum analogue of
randomized logspace, i.e., $\mbox{SL}\subseteq\mbox{QR}_{H}\mbox{L}$.
\end{abstract}


\section{Introduction}
\label{sec:introduction}

This paper addresses the problem of space-efficient quantum simulations of
probabilistic computations.
Although quantum Turing machines are known to be at least as powerful as
probabilistic Turing machines with respect to polynomial-time, bounded-error
computations \cite{BernsteinV97}, known methods for simulating bounded-error
probabilistic computations with quantum machines are either very
space-inefficient (as in the case of \cite{BernsteinV97}) or very
time-inefficient (as in the case of \cite{Watrous98c}) in the worst case.
The apparent difficulty in simulating probabilistic computations with
space-bounded quantum machines by means of the most straightforward technique
(i.e., directly simulating coin-flips with appropriately defined quantum
transformations) lies in the problem of reusing the space required for each
coin-flip, of which there may be a number exponential in the space-bound.

We prove in this paper that quantum Turing machines can simulate a limited
class of random processes---random walks on regular, undirected graphs---in a
time-efficient and space-efficient manner.
A random walk on a regular, undirected graph $G = (V,E)$ of degree $d$ is a
Markov chain defined as follows: the states of the Markov chain correspond to
the vertices of $G$, and the transition probability from vertex $u$ to vertex
$v$ is defined to be $1/d$ in case $v$ is adjacent to $u$, and zero otherwise.

The study of random walks has had a number of interesting applications in
complexity theory.
From the perspective of this paper, the most important such application is due
to Aleliunas, Karp, Lipton, Lov\'asz and Rackoff \cite{AleliunasK+79}, who
used random walks to show that the undirected graph connectivity problem can
be solved in $\mathrm{R}_{H}\mathrm{L}$ (sometimes written
$\mathrm{RL}^{\mathit{poly}}$ or simply $\mathrm{RL}$).
Since this problem is complete for symmetric logspace (SL) with respect to
logspace reductions \cite{LewisP82}, the relation
$\mbox{SL}\subseteq\mbox{R}_{H}\mbox{L}$ follows.
The most space-efficient deterministic algorithm for USTCON requires space
$O((\log n)^{4/3})$ \cite{ArmoniT+97}.
We focus on the variant of this problem in which the graph in question is
regular of a fixed degree~$d$:\vspace{1mm}

\begin{center}
\begin{minipage}{14cm}
\begin{center}
\underline{$d$-Regular Undirected Graph Connectivity (d-USTCON)}
\end{center}
\begin{tabular}{lp{11cm}}
Instance: & A regular, undirected graph $G=(V,E)$ of degree $d$ and
$s,t\in V$.\\
Question: & Are $s$ and $t$ connected in $G$?
\end{tabular}
\end{minipage}
\end{center}
\noindent
For $d\geq 3$, d-USTCON is SL-complete, as a straightforward reduction
shows $\mbox{USTCON}\leq_{m}^{\log}\mbox{d-USTCON}$.

By considering suitable quantum variants of random walks on graphs we prove
$\mbox{d-USTCON}\in\mbox{QR}_{H}\mbox{L}$, which is the quantum analogue of
$\mathrm{R}_{H}\mathrm{L}$.
This is done in two steps.
First we show $\mbox{d-USTCON}$ can be solved with one-sided error by
logspace quantum Turing machines that halt absolutely, but which have
considerably worse acceptance probability than 1/2 for positive instances.
We then demonstrate that $\mbox{QR}_{H}\mbox{L}$ is sufficiently robust with
respect to acceptance probabilities to yield
$\mbox{d-USTCON}\in\mbox{QR}_{H}\mbox{L}$.
This implies the following containment.
\begin{theorem}
\label{theorem:SL_in_QRL}
$\mathrm{SL}\subseteq\mathrm{QR}_{H}\mathrm{L}$.
\end{theorem}

Symmetric logspace is closed under complementation \cite{NisanT95}, which,
together with Theorem~\ref{theorem:SL_in_QRL}, implies
$\mbox{SL}\subseteq\mbox{QR}_{H}\mbox{L}\cap\mbox{co-QR}_{H}\mbox{L}=:
\mbox{ZQ}_{H}\mbox{L}$.

The remainder of this paper has the following organization.
First, in Section~\ref{sec:space-bounded_QTMs} we review relevant facts
concerning space-bounded quantum computation.
In Section~\ref{sec:Quantum_operators}, we define a number of quantum
operators and prove a lemma regarding these operators that will be useful in
Section~\ref{sec:QTM_construction}, which contains the construction of
quantum Turing machines for simulating classical random walks on $d$-regular
graphs.
In Section~\ref{sec:amplification}, we address the issue of robustness of
$\mbox{QR}_{H}\mbox{L}$ that, along with the machine constructed in 
Section~\ref{sec:QTM_construction}, allows us to deduce the main theorem.
Section~\ref{sec:conclusion} contains some concluding remarks.


\section{Space-bounded quantum Turing machines}
\label{sec:space-bounded_QTMs}

We begin by briefly reviewing some relevant facts concerning space-bounded
quantum computation; for further information see \cite{Watrous98c}.
For background on quantum computation more generally, we refer the reader to
\cite{BernsteinV97} and \cite{Berthiaume97}, and for classical space-bounded
computation see \cite{Saks96}.

The model of computation we use is the quantum Turing machine (QTM).
Our QTMs have three tapes: a read-only input tape, a work tape, and a write
only output tape.
The input and work tape alphabets are denoted $\Sigma$ and $\Gamma$,
respectively, and the output is assumed to be in binary.
Since our attention is restricted to decision problems, the only part of the
output we care about is the first output bit: 1 indicates acceptance and 0
indicates rejection.

The behavior of a QTM is determined by a transition function, along with an
observation of the output tape that is assumed to take place after each
computation step.
The computation continues as long as no output symbols are observed, and
acceptance or rejection is determined (in a probabilistic sense) by the
first output bit observed.
There are strict conditions the transition function of a QTM must satisfy, as
the evolution between observations must correspond to a norm-preserving
operator on the Hilbert space spanned by classical configurations of the
machine---see \cite{BernsteinV97, Watrous98c} for further discussion.

A QTM $M$ runs in logspace if there exists a function $f(n) = O(\log n)$ such
that, for every input $x$, the position of the work tape head of $M$ is never
outside the range $[-f(|x|),f(|x|)]$ with nonzero amplitude during its
computation on input $x$.

A QTM $M$ halts absolutely if, for each input $x$, there exists $t = t(x)$
such that the probability that $M$ halts (i.e., a 0 or a 1 is observed written
to the output tape) during the first $t$ steps of its computation on $x$ is 1.
If a QTM $M$ halts absolutely, it must do so in time at most exponential
in its space bound.
In particular, a logspace QTM that halts absolutely necessarily runs in
polynomial time.

The class $\mathrm{QR}_{H}\mathrm{L}$ consists of all languages $A$ for
which there exists a QTM $M$ that, on each input $x$, runs in logspace, halts
absolutely, and satisfies the following.
\begin{mylist}{8mm}
\item If $x\in A$, then $M$ accepts $x$ with probability at least 1/2.
\item If $x\not\in A$, then $M$ accepts $x$ with probability 0.
\end{mylist}
Substituting PTM for QTM in this definition yields the class
$\mathrm{R}_{H}\mathrm{L}$.
It is not currently known if $\mathrm{QR}_{H}\mathrm{L}$ and
$\mathrm{R}_{H}\mathrm{L}$ are different, nor if one is contained in the other.
In Section~\ref{sec:amplification} we show that the 1/2 in the above definition
for $\mathrm{QR}_{H}\mathrm{L}$ may be replaced by any function $f(|x|)$
satisfying $f(|x|)\geq 1/g(|x|)$ and $f(|x|)\leq 1-2^{-g(|x|)}$ for some
polynomial $g(|x|)>0$.

The quantum Turing machines we construct will be described using pseudo-code
in a manner typical for classical Turing machine descriptions.
Computations will be composed of transformations of two types: {\em quantum
transformations} and {\em reversible transformations} (both necessarily
inducing norm-preserving operators on the associated Hilbert space).
Quantum transformations will consist of a single step, so it will be trivial
to argue that each quantum transformation can be performed as claimed.
For reversible transformations, we rely on the result of Lange, McKenzie and
Tapp \cite{LangeM+97}, which implies that any logspace deterministic
computation can be simulated reversibly in logspace.
However, because the interference patterns produced by a given QTM depend
greatly upon the precise lengths of the various computation paths comprising
that machine's computation, we must take care to insure that these lengths are
predictable in order to correctly analyze the computation.
In the remainder of this section, we discuss reversible transformations
somewhat more formally, and state a theorem based on the main result of
\cite{LangeM+97} that will simplify this task.

For a given space-bound $f$ and work tape alphabet $\Gamma$, define
$W_{f(|x|)}(\Gamma)$ to be the set of all mappings of the form
$w:\Zed\rightarrow\Gamma$ taking the value \# (blank) outside the interval
$[-f(|x|),f(|x|)]$ (i.e., those mappings representing the possible contents of
the work tape of a machine on input $x$ having work tape alphabet $\Gamma$ and
running in space $f$).
By a {\em reversible transformation}, we mean a one-to-one and onto mapping of
the form $\Phi:W_{f(|x|)}(\Gamma)\rightarrow W_{f(|x|)}(\Gamma)$ for some $f$,
$x$ and $\Gamma$.
For a given machine $M$ having internal state set $Q$ and work tape alphabet
a superset of $\Gamma$, define $c(q,w)$ to be that configuration of $M$ for
which the work tape contents are described by $w$, the input and work tape
heads are scanning the squares indexed by 0, the internal state is $q$, and no
output has been written to the output tape.
Now, we say that a deterministic Turing machine $M$ on input $x$ performs
the transformation $\Phi$ on $W_{f(|x|)}(\Gamma)$ if the following conditions
are satisfied:
\begin{mylist}{5.5mm}
\item[1.] The work tape alphabet of $M$ is a superset of $\Gamma$.
\item[2.] The state set of $M$ includes two distinguished states $q_{0}$ and
$q_{f}$ (the initial state and final state).
\item[3.] If $M$ on input $x$ is placed in a configuration $c(q_{0},w)$ for
any $w\in W_{f(|x|)}(\Gamma)$, then there exists some positive integer
$t = t(x,w)$ such that if $M$ is run for precisely $t$ steps, it will then be
in configuration $c(q_{f},\Phi(w))$.
Furthermore, at no time prior to step number $t$ is the internal state of $M$
equal to $q_{f}$.
\end{mylist}
Naturally, we say that $t$ is the number of steps required for $M$ on $x$ to
perform $\Phi$.
If the work tape head of $M$ never leaves the region indexed by numbers in the
range $[-g(|x|),g(|x|)]$ during this process, we say that $M$ on $x$ performs
transformation $\Phi$ in space $g$.

\begin{theorem}
\label{theorem:LMT+uniform}
Let $f(n) = O(\log n)$ and let $M$ be a deterministic Turing machine that, on
each input $x$, performs reversible transformation $\Phi_{x}$ on
$W_{f(|x|)}(\Gamma)$ in space $O(\log |x|)$.
Then there exists a reversible Turing machine $M^{\prime}$ that, on each input
$x$, performs $\Phi_{x}$ on $W_{f(|x|)}(\Gamma)$ in space $O(\log |x|)$.
Furthermore, the number of steps required for $M^{\prime}$ to perform
$\Phi_{x}$ depends only on $x$ and not on the particular argument of
$\Phi_{x}$.
\end{theorem}
This theorem is based on a result due to Lange, McKenzie and Tapp
\cite{LangeM+97}, with added consideration payed to the number of steps
required for transformations.
See \cite{Watrous98c} (alternately \cite{Watrous98b}), along with
\cite{LangeM+97} for a proof.


\section{Quantum operators}
\label{sec:Quantum_operators}

In this section we define some operators and prove a key lemma that will be
used in the analysis of the machines in the next section.

Throughout this subsection, assume $G = (V,E)$ is an undirected, regular graph
of degree $d$ that is not necessarily connected.
The Hilbert space upon which the operators we define act is
$\mathcal{H} = \ell_{2}(V\times V)$, i.e., the classical states of our space
consists of all ordered pairs of vertices of $G$.
Let $n = |V|$, $m = |E|$, and for each $u\in V$ define
$S(u) = \{v\in V:\{u,v\}\in E\}$ and $B(u) = S(u)\cup\{u\}$.
Each operator we consider is linear: we define the action of operators on the
basis $\{\ket{u,v}:\,u,v\in V\}$ and extend to $\mathcal{H}$ by linearity.

First, define $F$ as follows:
\[
F \ket{u,v} = \left\{
\begin{array}{ll}
\displaystyle
\ket{u,v}-\frac{2}{d+1}\sum_{v^{\prime}\in B(u)}\smallket{u,v^{\prime}}
& v\in B(u)\\[2mm]
\ket{u,v} & v\not\in B(u).
\end{array}
\right.
\]
We now verify that $F$ is both unitary and hermitian.
Define
\[
\ket{\psi_{u}} = \frac{1}{\sqrt{d+1}}\sum_{v\in B_{u}}\ket{u,v}
\]
for each $u\in V$.
Note that $\{\ket{\psi_{u}}:u\in V\}$ is an orthonormal set.
We may rewrite $F$ as follows:
\begin{eqnarray*}
F & = & \sum_{u\in V}\sum_{v\in B(u)}
\left(\ket{u,v} - \frac{2}{d+1}\sum_{v^{\prime}\in B(u)}\smallket{u,v^{\prime}}
\right)\bra{u,v} + \sum_{u\in V}\sum_{v\not\in B(u)}\ket{u,v}\bra{u,v}\\
& = & \sum_{u,v\in V}\ket{u,v}\bra{u,v} - \frac{2}{d+1}\sum_{u\in V}
\left(\sum_{v,v^{\prime}\in B(u)}\smallket{u,v^{\prime}}\bra{u,v}\right)\\
& = & I - 2\sum_{u\in V}\ket{\psi_{u}}\bra{\psi_{u}}.
\end{eqnarray*}
Consequently, each vector $\ket{\psi_{u}}$ is an eigenvector of $F$ with
eigenvalue $-1$, and every vector orthogonal to
$\left\{\ket{\psi_{u}}:u\in V\right\}$ is an eigenvector of $F$ with
eigenvalue $1$.
From this it follows that $F$ is both unitary and hermitian:
$F = F^{\dagger} = F^{-1}$.
The operator $F$ is related to the operator $D$ defined on
$\ell_{2}(\{0,\ldots,d\})$ as follows:
\[
D\ket{a} = \ket{a} - \frac{2}{d+1}\sum_{b=0}^{d}\ket{b}.
\]
Up to a sign change, this is the ``diffusion'' operator used in Grover's
searching technique \cite{Grover96}.

Next, define $X$ as follows.
\[
X = \sum_{u,v\in V}\ket{v,u}\bra{u,v}.
\]
The operator $X$ simply exchanges the vertices $u$ and $v$.
Clearly $X = X^{\dagger} = X^{-1}$; $X$ is unitary and hermitian.

Finally, define a projection $P$ on $\mathcal{H}$ as
\[
P = \sum_{u\in V}\ket{u,u}\bra{u,u}.
\]

\begin{lemma}
\label{lemma:1}
Let $G = (V,E)$ be a regular graph of degree $d\geq 2$, let $F$, $X$ and $P$
be as defined above, define $Q = P\,F\,X\,F\,P$, and let
$k\geq\frac{d(d+1)^{2}n^{2}\log\,(1/\epsilon)}{8}$ for given $\epsilon>0$.
For each $u\in V$, let $G_{u} = (V_{u},E_{u})$ denote the connected component
of $G$ containing $u$, and write $n_{u} = |V_{u}|$.
Then for every $u\in V$ we have
\[
\left\|Q^{k}\ket{u,u}-\frac{1}{n_{u}}\sum_{v\in V_{u}}\ket{v,v}\right\|
<\epsilon.
\]
\end{lemma}
{\bf Proof.}
First, we note that
\begin{eqnarray}
Q\ket{u,u} & = & P\,F\,X\,F\,P\ket{u,u}\nonumber\\
& = & P\,F\,X\,F\ket{u,u}\nonumber\\
& = & P\,F\,X\left(\ket{u,u}-\frac{2}{d+1}\sum_{v\in B(u)}\ket{u,v}\right)
\nonumber\\
& = & P\,F\left(\ket{u,u}-\frac{2}{d+1}\sum_{v\in B(u)}\ket{v,u}\right)
\nonumber\\
& = & P\left(\ket{u,u}-\frac{2}{d+1}\sum_{v^{\prime}\in B(u)}
\smallket{u,v^{\prime}}- \frac{2}{d+1}\sum_{v\in B(u)}\left(\ket{v,u} - 
\frac{2}{d+1}\sum_{v^{\prime}\in B(v)}\smallket{v,v^{\prime}}\right)\right)
\nonumber\\
& = & \left(1-\frac{2}{d+1}\right)^{2}\ket{u,u}+
\left(\frac{2}{d+1}\right)^{2}\sum_{v\in S(u)}\ket{v,v}\label{eq:1}
\end{eqnarray}
for each $u\in V$, and clearly $Q\ket{u,v}=0$ for $u\not=v$.

For given $u\in V$ we have that $v\not\in V_{u}$ implies
$\opbracket{v,v}{Q^l}{u,u} = 0$ for $l=1$, and a simple induction shows that
this holds for any $l\geq 1$.
For each $u$, define $T_{u}$ to be a projection operator as follows:
\[
T_{u} = \sum_{v\in V_{u}}\ket{v,v}\bra{v,v}.
\]
Defining $Q_{u} = T_{u}QT_{u}$, we therefore have
$Q_{u}^{l}\ket{u,u}=Q^{l}\ket{u,u}$ for $t\geq 0$.
Note that $Q_{u}$ is hermitian:
$Q_{u}^{\dagger}=(T_{u}PFXFPT_{u})^{\dagger}=T_{u}PFXFPT_{u}=Q$, following
from the fact that $T_{u}$, $P$, $F$, and $X$ are hermitian.

Let $A$ denote the adjacency matrix of $G_{u}$ and let $f_{A}$ denote the
characteristic polynomial of $A$.
By (\ref{eq:1}), we determine that $f_{Q_{u}}$, the characteristic polynomial
of $Q_{u}$, satisfies
\begin{eqnarray*}
f_{Q_{u}}(z) & = &
z^{(n^{2}-n_{u})}\mbox{det}\left(z\,I_{n_{u}} -\left(1-\frac{2}{d+1}\right)^{2}
I_{n_{u}} - \left(\frac{2}{d+1}\right)^{2}\,A\right)\\
& = & z^{(n^{2}-n_{u})}\left(\frac{2}{d+1}\right)^{2n_{u}}
\mbox{det}\left(\frac{(d+1)^{2}z - (d-1)^{2}}{4}\,I - A\right)\\
& = & z^{(n^{2}-n_{u})}\left(\frac{2}{d+1}\right)^{2n_{u}}
f_{A}\left(\frac{(d+1)^{2}z - (d-1)^{2}}{4}\right).
\end{eqnarray*}
Letting $\lambda_{1}\geq\lambda_{2}\geq\cdots\geq\lambda_{n_{u}}$ be the
eigenvalues of $A$, we see that $Q_{u}$ has eigenvalues
\[
\mu_{j} = \frac{4\lambda_{j} + (d-1)^{2}}{(d+1)^{2}},
\]
for $j=1,\ldots,n_{u}$, as well as eigenvalues $\mu_{j} = 0$
for $j = n_{u}+1,\ldots, n^{2}$.
Note that the eigenvalues of $A$ (and hence the eigenvalues of $Q_{u}$)
are real since $A$ is symmetric.
Since $G_{u}$ is connected and regular of degree $d$, we have $\lambda_{1}=d$,
$\lambda_{j}<d$ for $j=2,\ldots,n_{u}$, and $\lambda_{n_{u}}\geq -d$
(see, e.g., \cite{Biggs74}, page 14).
Furthermore, it follows from \cite{LovaszW95} that
\[
\lambda_{j}\:\leq\: d - \frac{2}{dn_{u}^{2}},
\]
for $j = 2,\ldots, n_{u}$.
Hence $\mu_{1} = 1$, and
\[
\mu_{j}\in\left[1-\frac{8d}{(d+1)^{2}},\;1-\frac{8}{d(d+1)^{2}n_{u}^{2}}\right]
\]
for $j=2,\ldots,n_{u}$.
In particular, we have that $\mu_{2},\ldots,\mu_{n_{u}}$ are bounded in
absolute value by
\[
1 - \frac{8}{d(d+1)^{2}n_{u}^{2}}.
\]

Define
\[
\ket{\phi_{1}} = \frac{1}{\sqrt{n_{u}}}\sum_{u\in V_{u}}\ket{u,u}.
\]
It may be verified that $\ket{\phi_{1}}$ is an eigenvector of $Q_{u}$
corresponding to the eigenvalue $\mu_{1} = 1$.
As $Q_{u}$ is hermitian, we may choose eigenvectors
$\ket{\phi_{2}},\ldots,\ket{\phi_{n^{2}}}$ corresponding to eigenvalues
$\mu_{2},\ldots,\mu_{n^{2}}$ in such a way that
$\{\ket{\phi_{1}},\ldots,\ket{\phi_{n^{2}}}\}$ is an orthonormal basis of
$\mathcal{H}$.
Now, let $c_{j} = \bracket{\phi_{j}}{u,u}$ for $j=1,\ldots,n^{2}$.
We have $\ket{u,u} = \sum_{j=1}^{n^{2}}c_{j}\ket{\phi_{j}}$, and thus
\[
Q_{u}^{l}\ket{u,u} = \sum_{j=1}^{n_{u}}c_{j}\mu_{j}^{l}\ket{\phi_{j}},
\]
for $l\geq 1$.
Consequently,
\begin{equation}
\left\|	Q_{u}^{l}\ket{u,u}-\frac{1}{n_{u}}\sum_{v\in V_{u}}\ket{v,v}
\right\|^{2}
\:=\:\left\|\sum_{j=2}^{n_{u}}c_{j}\mu_{j}^{l}\ket{\phi_{j}}\right\|^{2}
\:=\: \sum_{j=2}^{n_{u}}|c_{j}|^{2}|\mu_{j}|^{2l}\:\leq\:
\left(1 - \frac{8}{d(d+1)^{2}n_{u}^{2}}\right)^{2l},
\label{eq:dist1}
\end{equation}
again for $l\geq 1$.

Now, since $k\geq\frac{d(d+1)^{2}n^{2}\log\,(1/\epsilon)}{8}$ for given
$\epsilon>0$, we have
\[
\left(1 - \frac{8}{d(d+1)^{2}n_{u}^{2}}\right)^{k} \leq
\left(1 - \frac{8}{d(d+1)^{2}n^{2}}\right)^{k} < \epsilon,
\]
for every $u$, following from the fact that $(1-1/x)^{x} < 1/e$ for $x\geq 1$.
Thus
\[
\left\|Q^{k}_{u}\ket{u,u} - \frac{1}{n_{u}}\sum_{v\in V_{u}}\ket{v,v}\right\|
< \epsilon
\]
follows by (\ref{eq:dist1}).
Since $Q^{k}\ket{u,u} = Q^{k}_{u}\ket{u,u}$, this completes the proof.
\qed

By taking $\epsilon = \frac{1}{2n}$ in Lemma~\ref{lemma:1}, we obtain the
following corollary.
\begin{cor}
Let $G = (V,E)$ be a regular graph of degree $d\geq 2$ with $s,t\in V$, let
$Q$ be as defined in Lemma~\ref{lemma:1}, and let
$k\geq\left\lceil d(d+1)^{2}n^{2}\log(2n)/8\right\rceil$.
If $s$ and $t$ are connected in $G$, then
\[
\left|\bra{\,t,t\,}Q^{k}\ket{s,s}\right|^{2}\:\geq\:\frac{1}{4n^2},
\]
and otherwise $\left|\bra{\,t,t\,}Q^{k}\ket{s,s}\right|^{2}=0$.
\end{cor}


\section{Quantum Turing machine construction and analysis}
\label{sec:QTM_construction}

We now construct, for each fixed degree $d\geq 2$, a logspace QTM solving
d-USTCON that operates with one-sided error and halts absolutely.
Although the QTMs we construct have somewhat poor probabilities of acceptance
for positive instances of d-USTCON, it will be demonstrated in the next
section that these machines may be modified to yield logspace QTMs for
d-USTCON having arbitrarily small one-sided error while still halting
absolutely.

\begin{lemma}
\label{lemma:2}
For $d\geq 2$, there exists a quantum Turing machine $M$ that runs in
logspace, halts absolutely, and operates as follows.
For any input encoding $(G,s,t)$, where $G = (V,E)$ is a regular, undirected
graph of degree $d$, $s,t\in V$, and $s$ is connected to $t$ in $G$, $M$
accepts with probability greater than $\frac{1}{4n^2}$, and for all other
inputs $M$ accepts with probability zero.
\end{lemma}
{\bf Proof.}
The work tape of $M$ will consist of four tracks, one for each of the
following variables: $u$, $v$, $b$ and $c$.
Each variable will contain an integer, with the exception of $v$, which will
store either an integer or a single symbol in the set $\{0,\ldots,d\}$.
Integers are assumed to be encoded as strings over the alphabet
$\{0^{\prime},1^{\prime}\}$, taken to be disjoint from $\{0,\ldots,d\}$.
We make the assumption that each integer has exactly one encoding and that
0 is encoded by the empty string.
Note that this implies $u$, $v$, $b$ and $c$ are all initially set to 0, as
the work tape initially contains only blanks.
Vertices of $G$ are assumed to be labeled by integers having length at most
logarithmic in the input size, and each vertex has a unique label.
When $u$ or $v$ contains an integer, this integer is to be interpreted as the
label of a vertex.

The execution of $M$ is described in Figure~\ref{fig:machine1}.
\begin{figure}[!ht]
\begin{flushleft}
\rule{6.79in}{0.2mm}
\end{flushleft}
\begin{center}
\begin{tabular}{lp{15cm}}
1. & Reject if the input does not encode $(G,s,t)$ for $G$ undirected and
     regular of degree $d$.\\
2. & Set $u = u + s$ and $v = v + s$. \\
3. & Loop with starting/stopping condition ``b=0'':\\
   & \begin{tabular}[t]{lp{13.5cm}}
     i.    & If $v\in B(u)$, replace $v$ with the symbol in $\{0,\ldots,d\}$
             corresponding to its index in $B(u)$ modulo $d+1$.\\
     ii.   & If $v\in\{0,\ldots,d\}$, perform transformation $D$ on $v$.\\
     iii   & Invert step i.\\
     iv.   & Exchange $u$ and $v$.\\
     v.    & If $v\in B(u)$, replace $v$ with the symbol in $\{0,\ldots,d\}$
             corresponding to its index in $B(u)$ modulo $d+1$.\\
     vi.   & If $v\in\{0,\ldots,d\}$, perform transformation $D$ on $v$.\\
     vii.  & Invert step v.\\
     viii. & In case $u\not=v$, increment $c$ modulo $d(d+1)^{2}n^{3} + 1$.\\
     ix.   & Increment $b$ modulo $d(d+1)^{2}n^{3}$.
     \end{tabular}\\
4. & If $c = 0$ and $u = t$, then {\em accept}, else {\em reject}.
\end{tabular}
\end{center}
\begin{flushleft}
\rule{6.79in}{0.2mm}
\end{flushleft}
\caption{Description of quantum Turing machine $M$ for Lemma~\ref{lemma:2}.}
\label{fig:machine1}
\end{figure}
For each of the steps in Figure~\ref{fig:machine1} we may define an
appropriate reversible or quantum transformation corresponding to the action
described.
Each transformation is to maintain the invariant that all tracks contain
strings having no embedded blanks and having leftmost symbol stored
in the work tape square indexed by 0.
The quantum transformations are steps ii and vi---these transformations
require a single step and involve only the symbol in square 0 of the track
corresponding to $v$.
The remaining transformations are reversible transformations---it is
straightforward to note that each such transformation may be performed by a
DTM running in space $O(\log n)$ in the manner described in
Section~\ref{sec:space-bounded_QTMs} for a suitable space-bound
$f(n) = O(\log n)$.
(It is for this reason that we increment $c$ modulo $d(d+1)^{2}n^{3} + 1$
instead of simply incrementing $c$ in step viii, although the same effect
results; each transformation must be defined on a bounded region of the work
tape).
We note that the quantity $d(d+1)^{2}n^{3}$ is somewhat arbitrary in steps
viii and ix---any quantity at least
$\left\lceil d(d+1)^{2}n^{2}\log (2n)/8\right\rceil$ suffices.
The loop may be implemented reversibly, in the manner described in
\cite{Watrous98c}.
By Theorem~\ref{theorem:LMT+uniform}, it follows that each reversible step in
Figure~\ref{fig:machine1} may be performed reversibly in logspace, requiring
time depending only on the input $(G,s,t)$ and not on the particular contents
of the work tape of $M$ when the step is performed.
Two consequences of this are (i) $M$ runs in logspace, and (ii) each step in
Figure~\ref{fig:machine1} may be viewed as requiring unit time, insofar as the
analysis of the machine is concerned.

Now let us analyze the computation of $M$ on a given input $(G,s,t)$.
When describing superpositions of $M$, we will restrict our attention to the
variables $u$, $v$, $b$ and $c$; since we will only care about superpositions
between the transformations described above, all other aspects of $M$
(specifically, tape head positions and internal state) are deterministic.
It will be most convenient to express such superpositions in terms of classical
states of the form $\ket{u,v}\ket{c}\ket{b}$ for $u,v\in V$, $c,b\in\Zed$,
which may be interpreted as being equivalent to classical states the form
$\ket{u,v,c,b}$.

Assume that $M$ does not reject during step 1, so that $G$ is indeed regular
of degree $d$ and undirected---otherwise $M$ of course functions as required.
After step 2 is performed, the superposition of $M$ is
$\ket{s,s}\ket{0}\ket{0}$.
Now the loop in step 3 is performed.
It may be verified that after one iteration of the loop, the superposition of
$M$ is $\left(Q\ket{s,s}\right)\ket{0}\ket{1} + \ket{\xi_{1,1}}\ket{1}\ket{1}$,
where $Q$ as defined in Section~\ref{sec:Quantum_operators} and
$\ket{\xi_{1,1}}$ is some vector (that we don't care about).
More generally, after $j< d(d+1)^{2}n^{3}$ iterations of the loop, the
superposition is
\[
\left(Q^{j}\ket{s,s}\right)\ket{0}\ket{j} +
\sum_{c\geq 1}\ket{\xi_{c,j}}\ket{c}\ket{j},
\]
and after $k = d(d+1)^{2}n^{3}$ iterations, the superposition is
\[
\left(Q^{k}\ket{s,s}\right)\ket{0}\ket{0} +
\sum_{c\geq 1}\ket{\xi_{c,0}}\ket{c}\ket{0}.
\]
At this point, the loop terminates, so that in step 4 the probability of
accepting is $\left|\bra{t,t}Q^{k}\ket{s,s}\right|^{2}$.
By Lemma~\ref{lemma:1}, we conclude that $M$ accepts $(G,s,t)$ with
probability at least $\frac{1}{4n^2}$ in case $s$ is connected to $t$, and
probability 0 otherwise.
\qed


\section{Amplification of Acceptance Probability}
\label{sec:amplification}

It is well-known that $\mbox{R}_{H}\mbox{L}$ is robust with respect to the
probability with which positive instances are accepted: the $1/2$ in the
definition of $\mbox{R}_{H}\mbox{L}$ may be replaced by any function $f(|x|)$
satisfying $f(|x|) \geq 1/g(|x|)$ and $f(|x|) \leq 1 - 2^{-g(|x|)}$
for $g(|x|) > 0$ a polynomial.
It is not immediate that an analogous fact holds for $\mbox{QR}_{H}\mbox{L}$;
repeated simulation a given QTM computation requires that the simulated
machine be in its initial configuration at the start of each simulation, but
resetting this machine to its initial configuration constitutes an
irreversible action that cannot be performed by the quantum machine performing
the simulation.
It is, however, not difficult to show that the analogous fact for
$\mbox{QR}_{H}\mbox{L}$ does hold by using a method described in
\cite{Watrous98c} for showing related facts regarding non-halting
versions of space-bounded classes.
Since the proof in the present case is similar, we will just sketch the proof.

\begin{lemma}
\label{lemma:amplify}
Let $M$ be a QTM that runs in logspace, halts absolutely, and accepts each
input $x$ with probability $p(x)$.
Then for any polynomial $f$, there exists a QTM $M_{f}$ that runs in
logspace, halts absolutely, and accepts each input $x$ with probability
\[
1 - (1 - p(x))(1 - 2 p(x))^{2f(|x|)}.
\]
\end{lemma}
{\bf Proof.} [Sketch]
Given $M$ and $f$ as in the statement of the theorem, we let $M_{f}$ be a
quantum Turing machine functioning as described in Figure~\ref{fig:amplify}.
\begin{figure}[!ht]
\begin{flushleft}
\rule{6.79in}{0.2mm}
\end{flushleft}
\begin{center}
\begin{tabular}{lp{15cm}}
1. & Repeat the following $f(|x|)+1$ times:\\
   & \begin{tabular}[t]{lp{13.5cm}}
     i.    & Simulate the computation of $M$ on $x$.\\
     ii.   & Accept if $M$ accepts $x$.\\
     iii.  & Invert step i.\\
     iv.   & If the current configuration of $M$ is not the initial
             configuration, multiply the current amplitude by -1.
     \end{tabular}\\
2. & Reject.
\end{tabular}
\end{center}
\begin{flushleft}
\rule{6.79in}{0.2mm}
\end{flushleft}
\caption{Description of quantum Turing machine $M_{f}$ for
Lemma~\ref{lemma:amplify}.}
\label{fig:amplify}
\end{figure}
We may assume without loss of generality that $M$ halts after precisely
$g(|x|)$ steps along all computation paths for $g$ bounded by a polynomial
and computable in logspace \cite{Watrous98c}, from which it follows that we
may take $M_{f}$ to run in logspace and halt absolutely.

During the simulation of $M$, $M_{f}$ will store an encoding of some
configuration of $M$ (including the first square on the output tape of $M$) on
its work tape.
Let us denote by $E$ an operator corresponding to performing step i on the
state of $M$ stored by $M_{f}$; since $M$ does not produce output during the
simulation, we may take $E$ to be unitary.
Initially, the state of $M$ represented by $M_{f}$ is $\ket{c_{0}}$, for
$c_{0}$ the initial configuration of $M$.
After performing step i for the first time, this state is mapped to
$E\ket{c_{0}} = \ket{\psi}$.
Write $\ket{\psi} = \ket{\psi_{acc}} + \smallket{\psi_{acc}^{\perp}}$, where
$\ket{\psi_{acc}}$ denotes the projection of $\ket{\psi}$ onto the space
spanned by accepting configurations of $M$ and $\smallket{\psi_{acc}^{\perp}}$
is orthogonal to $\ket{\psi_{acc}}$.
During step ii, $M_{f}$ accepts $x$ with probability
$p(x) = \|\ket{\psi_{acc}}\|^{2}$, and otherwise the superposition of $M_{f}$
collapses so that $\smallket{\psi_{acc}^{\perp}}$ is the state of $M$
represented.

Now we consider the sequence of steps iii, iv, i, ii, which are at this point
performed $f(|x|)$ times.
It can be checked that each iteration of the sequence of steps ii, iv, i has
the effect of mapping $\smallket{\psi_{acc}^{\perp}}$ to
$(2 - 2p(x))\ket{\psi_{acc}} + (1 - 2p(x))\smallket{\psi_{acc}^{\perp}}$
(following from the fact that
$\opbracket{c_{0}}{E^{\dagger}}{\psi_{acc}}=p(x)$).
The $j$-th time step i is performed, for $j\geq 2$, thus results in acceptance
with probability $p(x)(1 - 2p(x))^{2(j-2)}(2 - 2p(x))^2$.
We therefore have that $M_{f}$ accepts with probability
\[
p(x) + \sum_{j=2}^{f(|x|)+1}p(x)(1 - 2p(x))^{2(j-2)}(2 - 2p(x))^2
\:=\: 1 - (1-p(x))(1 - 2p(x))^{2f(|x|)}
\]
as claimed.
\qed

By Lemma~\ref{lemma:2} and Lemma~\ref{lemma:amplify}, we have
$\mbox{d-USTCON}\in\mbox{QR}_{H}\mbox{L}$.

Theorem~\ref{theorem:SL_in_QRL} now follows in straightforward fashion,
relying again on Theorem~\ref{theorem:LMT+uniform}; given a particular
language $A\in\mbox{SL}$ we have a logspace many-one reduction to d-USTCON,
and we may replace various reversible transformations of our machine for
d-USTCON with appropriately defined reversible transformations based on
compositions of the reduction with the replaced transformation.
Details will appear in the final version of this paper.


\section{Concluding Remarks}
\label{sec:conclusion}

In this paper we have shown that logspace quantum Turing machines can
simulate a limited class of probabilistic computations in a time-efficient
manner.
This leaves open the question of whether probabilistic computations can be
simulated efficiently by space-bounded quantum machines in general (e.g., is
$\mbox{R}_{H}\mbox{L}$ contained in $\mbox{QR}_{H}\mbox{L}$?)
We hope our techniques will provide insight into this problem.

We have defined in this paper quantum processes that attempt to mimic
classical random walks on graphs.
There are a number of ways in which to define {\em quantum walks on graphs}
having properties quite different from classical random walks.
It may be interesting to consider possible applications of such processes to
quantum complexity theory.



\end{document}